\documentclass[pra,twocolumn,showpacs,a4paper]{revtex4}
\usepackage{epsfig}
\usepackage{amsmath}
\usepackage{graphicx}
\usepackage{bm}
\usepackage{ulem}
\usepackage{amsfonts}
\usepackage{amssymb}
\usepackage{amsthm}
\theoremstyle{plain}
\newtheorem{thm}{Theorem}
\newtheorem{cor}{Corollary}
\newtheorem{lem}{Lemma}
\theoremstyle{definition}
\newtheorem{defn}{Definition}
\theoremstyle{remark}

\begin{document}
\title{Optimal fidelity of teleportation of coherent states and entanglement}
\author{A. Mari and D. Vitali}
\affiliation{Dipartimento di Fisica, Universit\`{a} di Camerino, I-62032 Camerino (MC), Italy}

\begin{abstract}
We study the Braunstein-Kimble protocol for the continuous variable teleportation of a coherent state. We determine lower and upper bounds for
the optimal fidelity of teleportation, maximized over all local Gaussian operations for a given entanglement of the two-mode Gaussian state
shared by the sender (Alice) and the receiver (Bob). We also determine the optimal local transformations at Alice and Bob sites and the
corresponding maximum fidelity when one restricts to local trace-preserving Gaussian completely positive maps.
\end{abstract}


\maketitle

\section{Introduction}

Quantum teleportation \cite{Tel93} is the transfer of an unknown quantum state from a sender (Alice) to a receiver (Bob) by means of the
entanglement shared by the two parties and appropriate classical communication. Bob recovers an exact copy of the state teleported to him by
Alice only if the quantum channel is the ideal maximally entangled state, which however, in the case of continuous variables (CV), is an
unphysical infinitely squeezed state \cite{vaidman}. Nevertheless, by considering the finite quantum correlations between the quadratures in a
two-mode squeezed state, Braunstein and Kimble \cite{braKim98} proposed a realistic protocol employing a beam splitter and homodyne
measurements, which approaches perfect teleportation in the limit of infinite degree of squeezing. This protocol (and its various extensions)
has been then implemented by various groups \cite{telepexper}. The success of a teleportation experiment is quantified by the \textit{fidelity}
which, in the case of a pure input state $|\psi_{in}  \rangle  $, is given by $\mathcal F =\langle \psi_{in} | \rho_{out}|\psi_{in}\rangle $,
where $\rho_{out}$ denotes the output state of the protocol, and coincides with the probability of finding the input state $|\psi_{in}\rangle $
at the output.

The relation between the fidelity of CV teleportation $\mathcal F$ and the entanglement shared by the distant parties is nontrivial and it has
been investigated in various papers \cite{braKim00,ade-illu05}. In fact, entanglement is the key resource that allows to beat any classical
strategy for transmitting quantum states. One can say that genuine quantum teleportation has been performed only if $\mathcal F >
\mathcal{F}_{cl}$, where $\mathcal{F}_{cl}$ is the classical fidelity threshold achievable by two cheating parties who can perform arbitrary
local operations and classical communication (LOCC) but are not able to share entanglement, nor to directly transmit quantum systems
\cite{braKim00}. This means that $\mathcal F > \mathcal{F}_{cl}$ is a sufficient condition for entanglement between Alice and Bob; it is not a
necessary condition because one can have lower-than-classical fidelities even using an entangled state. This is due to the fact that the
Braunstein-Kimble protocol chooses specific combinations of quadratures, and therefore $\mathcal F$, differently from entanglement, is not a
local symplectic invariant: if the protocol employs combinations of quadratures which are inappropriate for the given shared entangled state,
the resulting $\mathcal F$ is very low. The difference between of teleportation fidelity and entanglement opens two problems: i) the
determination of the classical fidelity threshold $\mathcal{F}_{cl}$ for a given class of CV input states; ii) the optimization of the
teleportation fidelity, for the chosen class of input states, over all possible LOCC strategies.
Determining $\mathcal{F}_{cl}$ is a nontrivial quantum estimation problem which has been solved only for few classes of states: i) input
coherent states with completely unknown amplitude ($\mathcal{F}_{cl}=1/2$) \cite{braKim00}; ii) pure squeezed states with zero displacement and
completely unknown degree of squeezing at a given phase (see Ref.~\cite{ade-chi08} which obtained $\mathcal{F}_{cl} = 0.81$ by considering LOCC
strategies in which one is only allowed to prepare squeezed thermal states); iii) pure squeezed states with completely unknown displacement and
orientation in phase space but fixed degree of squeezing $s$ ($\mathcal{F}_{cl}=\sqrt{s}/(1+s)$) \cite{owari}. Here we restrict to input
coherent states, which represent the basic resource for many quantum communication schemes \cite{cryptocv,pirnatphys}.

The improvement of the teleportation of coherent states by means of local operations and its relation with the entanglement of the shared
entangled state has been already discussed in a number of papers \cite{Bow,kim,fiu02,ade-illu05}. Ref.~\cite{Bow} showed that in some cases the
fidelity of teleportation may be improved by local squeezing transformations, while Ref.~\cite{kim} showed that in the case of a shared
asymmetric mixed entangled resource, teleportation fidelity can be improved even by a local \textit{noisy} operation.
Ref.~\cite{fiu02} then considered the class of local trace-preserving Gaussian completely positive (TGCP) maps (those performed by first adding
ancillary systems in Gaussian states, then performing unitary Gaussian transformations on the whole system, and finally discarding the
ancillas), and maximized the fidelity over this class of operations.

Ref.~\cite{fiu02} confirmed that the best local TGCP map maybe a noisy one, i.e., that teleportation fidelity can be increased even by
decreasing the entanglement and increasing the noise of the shared entangled state. Ref.~\cite{fiu02}, however, did not discuss the relationship
between entanglement and the optimal fidelity $\mathcal{F}_{opt}$. Ref.~\cite{ade-illu05} instead found this relationship, but only for a
subclass of symmetric Gaussian entangled state shared by Alice and Bob: for this class it is $\mathcal{F}_{opt}=\left(1+\nu\right)^{-1}$, where
$\nu$ is the lowest symplectic eigenvalue of the partial transposed (PT) state. The parameter $\nu$ provides a quantitative characterization of
CV entanglement, because the logarithmic negativity $E_{\mathcal{N}}$ is related to $\nu$ by $E_{\mathcal{N}}=\max [0,-\ln \nu]$ \cite{logneg}.

The problem of finding the optimal LOCC strategy is non-trivial because the set of operations that Alice and Bob can adopt is very large. In
fact, apart from local TGCP maps, they can adopt two further options: i) use \textit{non-trace preserving} Gaussian operations in which some
ancillary mode is subject to Gaussian measurement, i.e., projected onto a Gaussian state, rather than discarded \cite{giedkemap}; ii) use local
non-Gaussian operations (either with measurement on ancillas or not), i.e., those involving interactions which are non-quadratic in the
canonical coordinates. The first class of maps, together with TGCP maps, forms the most general class of Gaussian completely positive (GCP)
operations, capable of preserving the Gaussian nature of the state shared by Alice and Bob. Non-Gaussian operations instead will transform the
initial Gaussian bipartite state of Alice and Bob into a non-Gaussian one, and they can also increase the fidelity of teleportation in some
cases \cite{nonGtelep}.

In this paper we generalize in various directions the results of Refs.~\cite{fiu02,ade-illu05}. We show that if Alice and Bob share a bipartite
Gaussian state with a given $\nu$ and one restricts to local GCP maps which preserve such a Gaussian nature, the optimized fidelity always
satisfies
\begin{equation}\label{lowupp}
    \frac{1+\nu}{1+3\nu} \leq \mathcal{F}_{opt}\leq \frac{1}{1+\nu}.
\end{equation}
We also show that the upper bound is reached iff Alice and Bob share a symmetric entangled state. Moreover we determine the optimal local
transformations at Alice and Bob sites and the corresponding value of $\mathcal{F}_{opt}$ as a function of the symplectic invariants of the
shared CV entangled state when one restricts to local TGCP maps.

The paper is organized as follows. In Sec.~II we provide the basic definitions of the problem, in Sec.~III we prove and discuss the lower and
upper bounds for the optimal teleportation fidelity for a given shared entanglement. In Sec.~IV we discuss the properties of the optimal local
map and derive its explicit form in the case of TGCP maps. Sec.~V is for concluding remarks.

\section{Definition of the problem}

The protocol for a perfect CV quantum teleportation based on ideal Einstein-Podolski-Rosen (EPR) correlations has been introduced by Vaidman
\cite{vaidman}, and then adapted to the finite correlations of a two-mode squeezed states by Braunstein and Kimble \cite {braKim98}. The idea
can be easily shown in the Heisenberg picture. We consider two CV systems, each described by a pair of conjugate dimensionless quadratures
$\hat{x}_{k}$ and $\hat{p}_{k}$ ($k=a,b$). Introducing the vector $\hat{\xi}^{T}\equiv (\hat{x}_{a},\hat{p}_{a},\hat{x} _{b},\hat{p}_{b})$, we
can write the canonical
commutation relations as $[\hat{\xi}_{l},\hat{\xi}_{m}]=i\mathcal{J}%
_{lm}$ ($l,m=1,...4$), where
\begin{equation}
\mathcal{J}\equiv J\oplus J,~J\equiv \left(
\begin{array}{cc}
0 & 1 \\
-1 & 0
\end{array}
\right) \text{ ,}  \label{sympl2}
\end{equation}
and $\oplus $ denotes the usual direct sum operation.

Alice and Bob share a bipartite CV state with EPR-like correlations, i.e., a state which can be considered as an approximate simultaneous
eigenstate of the combinations of quadratures $\hat{x}_{a}+\hat{x}_{b}$ and $\hat{p}_{a}-\hat{p}_{b}$, so that the variances of these two
combinations are both close to zero. Alice also possesses an unknown input state with quadratures $\hat{x}_{in},\hat{p}_{in}$ which she wants to
teleport to Bob. Alice mixes the input mode with her part of the entangled state via a balanced beam splitter and she carries out a homodyne
detection on each output mode, thereby measuring two commuting quadratures $\hat{x}_{+}=(\hat{x}_{a}+\hat{x}_{in})/\sqrt{2}$ and
$\hat{p}_{-}=(\hat{p}_{in}-\hat{p}_{a})/\sqrt{2}$. After receiving the measured values $x_{+}$ and $p_{-}$ from Alice, Bob uses this transmitted
classical information to perform a suitable conditional displacement on his own mode, $\hat{x}_{b}\longrightarrow \hat{x}_{b}^{\prime }\equiv
\hat{x}_{b}+\sqrt{2}x_{+}$, $\hat{p}_{b}\longrightarrow \hat{p}_{b}^{\prime }\equiv \hat{p}_{b}+\sqrt{2} p_{-}$.

If we assume ideal homodyne detectors on Alice site, and that the shared bipartite state is undisplaced (i.e., all mean values of Alice and Bob
quadratures vanish), the EPR-like correlations $\hat{x}_{a}\simeq - \hat{x}_{b}$ and $\hat{p}_{a}\simeq \hat{p}_{b}$, together with Bob
displacements, imply $\hat{x}_{b}^{\prime }\simeq \hat{x}_{in}$ and $\hat{p}_{b}^{\prime }\simeq \hat{p}_{in}$, i.e., Bob mode is described by a
pair of conjugate variables very close to those of the input mode. In the Schr\"odinger picture, this is equivalent to teleport the input state
to Bob with a fidelity very close to one.

We restrict to the case when the state shared by Alice and Bob $\rho_{ab}$ is Gaussian, where a compact expression of the resulting fidelity of
teleportation can be derived \cite{fiu02,pirrew,marian}. A bipartite CV state $\rho _{ab}$ is Gaussian if its Wigner characteristic function
$\Phi _{ab}( \vec{\xi})\equiv \mathrm{Tr}[\rho _{ab}\exp (-i\vec{\xi}^{T}\cdot\hat{\xi})]$ (where $\vec{\xi}^{T}=(x_{a},p_{a},x_{b},p_{b})$ is
the vector of phase-space variables corresponding to $\hat{\xi}^{T}$), is Gaussian, i.e., $\Phi _{ab}(\vec{\xi})=\exp
(-\vec{\xi}^{T}V\vec{\xi}/4+i \vec{d}^{T}\vec{\xi})$. We have assumed that Alice and Bob share a zero-displacement state, implying $ \vec{d}
=0$. Therefore $\rho _{ab}$ is fully characterized only by its correlation matrix (CM) $V$, whose generic element is defined as $V_{lm}\equiv
\langle \Delta
\hat{\xi}_{l}\Delta \hat{\xi} _{m}+\Delta \hat{\xi}_{m}\Delta \hat{\xi}%
_{l}\rangle $ where $\Delta \hat{ \xi}_{l}\equiv \hat{\xi}_{l}-\langle \hat{\xi}_{l}\rangle $. The CM satisfies the uncertainty principle
$V+i\mathcal{J}\geq 0 $ \cite{mukunda}, and can always be put in the block form
\begin{equation}
V\equiv \left(
\begin{array}{cc}
A & C \\
C^{T} & B
\end{array}
\right) \text{ ,}  \label{blocks}
\end{equation}
where $A,B,$ and $C$ are $2\times 2$ real matrices. Using characteristic functions, it is straightforward to prove \cite{fiu02,pirrew} that, if
the input state is a single-mode Gaussian state with CM $V_{in}$, the fidelity of teleportation is given by
\begin{equation}
\mathcal F=\frac{2}{\sqrt{\det{(2 V_{in}+N)}}},\label{fid}
\end{equation}
where
\begin{equation} \label{noise}
N=ZAZ+ZC+C^T Z+B,
\end{equation}
with $Z=\mathrm{diag}(1,-1)$ \cite{notedispl}. The $2\times 2$ matrix $N$ is semipositive definite, $N \geq 0$, it describes the noise added to
the teleported state, and it is equal to zero only in the ideal situation of perfect EPR correlations between Alice and Bob. As discussed in the
introduction, we shall restrict to the case of input coherent states, $V_{in}=I$, so that Eq.~(\ref{fid}) reduces to
\begin{equation}
\mathcal F=\frac{2}{\sqrt{4+2 \textrm{Tr} N+\det N}}\label{fidco}.
\end{equation}
The problem afforded in this paper, i.e., the maximization of the teleportation fidelity over all possible Gaussian LOCC strategies for a given
Alice-Bob entanglement, therefore means to determine the optimal local transformation of matrices $A$, $B$ and $C$ which makes $N$ as small as
possible.

As showed in \cite{marian}, using an unbalanced beam splitter is equivalent, for the teleportation protocol, to a squeezing operation by Alice.
Therefore the optimization over all Alice and Bob local operations includes also any eventual modification of the beam splitter used for the
joint homodyne measurement.

\section{Upper and lower bounds for the fidelity of teleportation}

\begin{figure}[tbh]
\centerline{\includegraphics[width=0.45\textwidth]{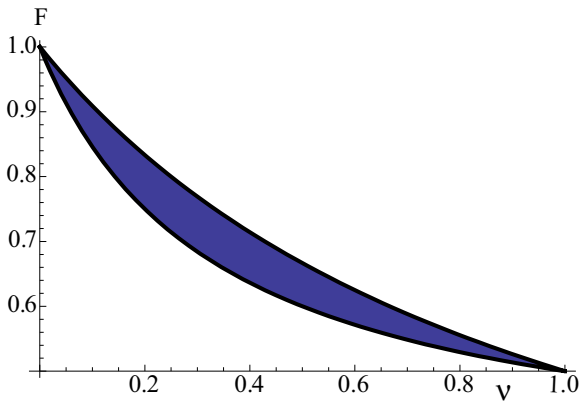}} \caption{(Color online) Plot of the upper and lower bounds (Eq.~(\ref{ubound})
and (\ref{lbound}) respectively) for the fidelity of teleportation of coherent states. The blue region is the allowed region in the $(\mathcal
F,\nu)$ plane.} \label{bounds}
\end{figure}

\begin{figure}[tbh]
\centerline{\includegraphics[width=0.45\textwidth]{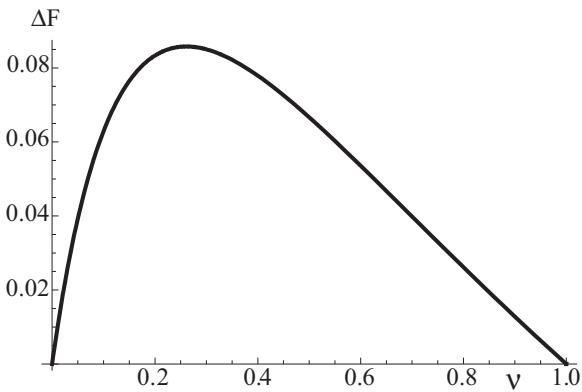}} \caption{Plot of the distance between the upper and lower bounds for the
teleportation fidelity versus the allowed values of $\nu$ for an entangled state between Alice and Bob. We see that the error $\Delta \mathcal
F$ under which we can estimate the maximum fidelity is less than $0.086$.} \label{error}
\end{figure}

In this section we prove Eq.~(\ref{lowupp}), i.e., the upper and lower bounds for the fidelity of teleportation for input coherent states. An
important preliminary result enabling us to derive the two bounds is the fact that the optimal noise matrix $N$ is very simple: in fact, the
maximum teleportation fidelity is obtained when $N$ is proportional to the $2\times 2$ identity matrix $I$. More precisely, we have the
following
\begin{lem}[\textbf{Optimal noise matrix}]
If $\omega_{opt}$ is an optimal local GCP map which gives the maximum of the fidelity $\mathcal F_{max}$ for the teleportation of a coherent
state, then the resulting noise matrix is a multiple of the identity, that is $N_{opt}=2n_{opt}I$. \label{lemN}
\end{lem}
\begin{proof}
First of all we observe that for a $2 \times 2$, symmetric and positive semidefinite matrix like $N$, the condition $N=2n_{opt}I$ is equivalent
to $\textrm{Tr} N=2\sqrt{\det N}$. Therefore we have to show that $\omega_{opt}$ is such that $\textrm{Tr}N_{opt}=2\sqrt{\det N_{opt}}$. We do
this by reductio ad absurdum supposing that $\omega_{opt}$ gives a noise matrix with $\textrm{Tr}N_{opt}>2\sqrt{\det N_{opt}}$. However, within
the class of local GCP maps, there exists a subclass of local symplectic (i.e., unitary Gaussian) maps realized by a generic symplectic $S_{b}$
on Bob mode, and the associated symplectic map $ S_{a}=Z S_{b}Z$ on Alice mode, which act as an \textit{effective} symplectic transformation on
$N_{opt}$, $N_{opt}'=S_b N_{opt} S_b^T$ (see Eq.~(\ref{noise})). We can always choose $S_b$ such that $N_{opt}'=\sqrt{\det N_{opt}}I$, for which
$\textrm{Tr}{N_{opt}'}=2 \sqrt{\det N_{opt}}<\textrm{Tr}{N_{opt}}$ while $\det N_{opt}'=\det N_{opt}$. However, we see from Eq.~(\ref{fidco}),
that this local symplectic operation \textit{increases} the teleportation fidelity, but this is absurd because we assumed from the beginning
that $N_{opt}$ is optimal.
\end{proof}
From this lemma and Eq.~(\ref{fidco}) we can therefore rewrite the optimal fidelity of teleportation in terms of the single positive parameter
$n_{opt}=\sqrt{\det N_{opt}}/2$ as
\begin{equation}
 \mathcal F_{opt} = \frac{1}{1+n_{opt}}.\label{fidn}
\end{equation}
We can now derive an upper bound for $\mathcal F_{opt}$ for a given entanglement of the state shared by Alice and Bob. We quantify such
entanglement in terms of the lowest partially transposed (PT) symplectic eigenvalue, $\nu$. Such a parameter cannot be improved (i.e.,
decreased) by local operations and therefore provides a quantitative characterization of CV entanglement \cite{logneg}. It is a local symplectic
invariant and it can be expressed in terms of the four local symplectic invariants $\det A$, $\det B$, $\det C$ and $\det V$ as
\begin{equation}
\nu = 2^{-1/2}\left[ \Sigma (V)- \left( \Sigma (V)^{2}-4\det V\right) ^{1/2}\right] ^{1/2},  \label{Sympl_eigenv}
\end{equation}
where $\Sigma (V)\equiv \det A+\det B-2\det C$.
\begin{thm}[\textbf{Upper bound}] For a given Gaussian bipartite state shared by Alice and Bob, with lowest PT symplectic eigenvalue
$\nu$, the fidelity of the teleportation of a coherent state is limited from above by
\begin{equation}
\mathcal F_{opt}\le \frac{1}{1+ \nu}. \label{ubound}
\end{equation}
\end{thm}
\begin{proof}
Let us suppose that we can achieve a larger fidelity $\mathcal F=1/(1+n_{opt})$ with $0<n_{opt}<\nu$. Alice can in principle have at her
disposal a two-mode squeezed state, with the usual correlation matrix
\begin{equation}
W=
\left(
\begin{array}{cc}
I\cosh r & -Z\sinh r \\
-Z\sinh r & I\cosh r \\
\end{array}
\right),
\end{equation}
($r$ is the squeezing parameter) and use this two-mode squeezed state, together with the bipartite state shared with Bob already optimized over
all local GCP maps, to implement a CV entanglement swapping protocol \cite{swap}. In fact, by mixing at a balanced beam splitter her mode of the
bipartite state shared with Bob and one part of the two-mode squeezed state, and performing homodyne measurements at the output, Bob mode gets
entangled with the remaining part of the two-mode squeezed state in Alice hands. Since the noise added to the teleported state is
$N_{opt}=2n_{opt}I$, it is straightforward to see that the two remaining modes are then described by the following CM
\begin{equation}
W_{swap}=
\left(
\begin{array}{cc}
I\cosh r & -Z\sinh r \\
-Z\sinh r & I[2n_{opt}+\cosh r] \\
\end{array}
\right).
\end{equation}
In other words, before entanglement swapping, Alice and Bob shared an entangled state with CM $V$ and entanglement characterized by $\nu$; after
entanglement swapping, they share a state with CM $W_{swap}$. In the limit of infinite squeezing the lowest PT symplectic eigenvalue of
$W_{swap}$ tends to $n_{opt}$, i.e., $\lim_{r\rightarrow \infty} \nu_{swap}=n_{opt}$. Since we supposed $n_{opt}< \nu$, this means that for a
sufficiently large squeezing parameter $r$, $\nu_{swap}<\nu$, i.e., Alice and Bob have increased their entanglement. However this is impossible
because we have employed only local operations~\cite{noterefe}. Therefore it must be $n_{opt} \geq \nu$.
\end{proof}
We complete the characterization of the optimal fidelity of teleportation in terms of the entanglement shared by the two distant parties by
providing also a lower bound for $\mathcal F_{opt}$, proving in this way the result of Eq.~(\ref{lowupp}).
\begin{thm}[\textbf{Lower bound}] For a given Gaussian bipartite state shared by Alice and Bob, with lowest PT symplectic eigenvalue
$\nu$, the fidelity of the teleportation of a coherent state is limited from below by
\begin{equation}
\mathcal F_{opt} \ge \frac{1+\nu}{1+3\nu}.\label{lbound}
\end{equation}
\end{thm}
\begin{proof}
From the definition of symplectic eigenvalue, one has that a $4 \times 4$ symplectic matrix $S$ exists which diagonalizes $\Lambda V \Lambda$
($\Lambda=\textrm{diag}(Z,I)$), i.e., the PT matrix of the CM $V$. This means $S \Lambda V \Lambda S^T=\textrm{diag}(\nu,\nu,\mu,\mu)$, where
$\mu$ is the largest PT symplectic eigenvalue. By writing $S$ in $2\times 2$ block form
\begin{equation}
S=
\left(
\begin{array}{cc}
 W_a & W_b \\
 W_c & W_d \\
\end{array}
\right),
\end{equation}
and rewriting the diagonalization condition for the upper $2 \times 2$ block only, one gets the following condition
\begin{equation}\label{relation}
W_a Z A Z W_a^T + W_a Z C W_b^T +W_b C^T Z W_a^T + W_b B W_b^T=\nu I.
\end{equation}
The symplectic transformation $S$ transforms the vector of quadratures $\hat{\xi}$ into $\hat{\xi}'=(x_a',y_a',x_b',y_b')^T=S \hat{\xi}$ and the
PT vector into $\hat{\xi}''=(x_a'',y_a'',x_b'',y_b'')^T=S\Lambda \hat{\xi}$. One has $[x_a',y_a']=i$, because commutation relation are preserved
by $S$, implying
\begin{equation} \label{det1}
\det W_a+\det W_b=1.
\end{equation}
The commutation relation is instead not preserved for the PT transformed quadratures, and introducing a real parameter $\epsilon$ such that
$[x_a'',y_a'']=i\epsilon$, we get another condition for the two upper blocks of $S$,
\begin{equation}
-\det W_a+\det W_b= \epsilon,
\end{equation}
which together with Eq.~(\ref{det1}), gives the parametrization
\begin{equation} \label{detsol}
 \det W_a=(1 -\epsilon)/2,\qquad \det W_b=(1 +\epsilon)/2.
\end{equation}
Now, since $\Delta x_a''^2= \Delta y_a''^2=\nu/2$, the Heisenberg uncertainty principle imposes that $|\epsilon|\le\nu$ and in particular for
every entangled state we have $|\epsilon|\le\nu <1$. This latter condition, together with Eq.~(\ref{detsol}), suggests an alternative
parametrization in terms of the angle $\theta=\arctan\sqrt{(1-\epsilon)/(1+\epsilon)}$ ($0 < \theta < \pi/2$),
\begin{equation} \label{det theta}
 \sqrt{\det W_a}=\sin\theta,\quad \sqrt{\det W_b}=\cos\theta .
\end{equation}
The $2 \times 2$ matrices $W_a$ and $W_b$ and the parameter $\theta$ allow to construct an appropriate local map which will lead us to derive a
lower bound for the fidelity. This local map is a TGCP map which, at the level of CM, acts as \cite{TGCP1,lin00,TGCP2}
\begin{equation}
V \rightarrow V'=SVS^T+G, \label{genTGCP}
\end{equation}
with $S$ and $G$ satisfying \begin{equation}\label{condiTGCP}
 G+i\mathcal J-iS\mathcal J S^T \geq 0.
\end{equation}
If the TGCP map is local, then $S=S_a
\oplus S_b$ and $G= G_a \oplus G_b$, with $G_k+i J-iS_k J S_k^T \geq 0$ ($k=a,b$).

The desired local TGCP map $\omega_\theta$ is defined in terms of $S_a$, $S_b$, $G_a$ and $G_b$ in the following way
\begin{subequations}
\label{tgcpmap}
\begin{eqnarray}
S_a & = & \left\{\begin{array}{cc}
 ZW_aZ \left[\cos\theta\right]^{-1} & 0 < \theta \leq \pi/4 \\
 ZW_aZ\left[\sin\theta\right]^{-1} & \pi/4 \leq \theta < \pi/2 , \\
\end{array}
\right.\\
S_b & = & \left\{\begin{array}{cc}
 W_b\left[\cos\theta\right]^{-1} & 0 < \theta \leq \pi/4 \\
 W_b\left[\sin\theta\right]^{-1} & \pi/4 \leq \theta < \pi/2,  \\
\end{array}
\right.\\
G_a & = & \left\{\begin{array}{cc}
 \left[1-\tan^2\theta\right]I & 0 < \theta \leq \pi/4 \\
 0 & \pi/4 \leq \theta < \pi/2 , \\
\end{array}
\right.\\
G_b & = & \left\{\begin{array}{cc}
 0 & 0 < \theta \leq \pi/4 \\
 \left[1-\cot^2\theta\right]I & \pi/4 \leq \theta < \pi/2 . \\
\end{array}
\right. 
\end{eqnarray}
\end{subequations}
By applying Eqs.~(\ref{noise}), (\ref{relation}) and (\ref{genTGCP}), one can see that this local TGCP map transforms the noise matrix $N$ into
a final matrix proportional to the identity, given by
\begin{eqnarray}
 N&=& [\nu/\cos^2\theta+ 1-\tan^2\theta]I,\quad 0 < \theta \leq \pi/4,\label{noiseb} \\
 N&=&[\nu/\sin^2\theta+1-\cot^2\theta]I,\quad \pi/4 \leq \theta < \pi/2.\label{noisea}
\end{eqnarray}
It is however convenient to come back to the parametrization in terms of $\epsilon$, which allows to express the final $N$ in a unique way, for
$0 < \theta < \pi/2$. In fact, from Eqs.~(\ref{noiseb})-(\ref{noisea}), one gets
\begin{equation}
N=2 \frac{\nu+|\epsilon|}{1+|\epsilon|}I,
\end{equation}
which, inserted into Eq.~(\ref{fidn}), yields
\begin{equation}
 \mathcal F=\frac{1+|\epsilon|}{1+\nu+2|\epsilon|}.\label{fideps}
\end{equation}
From the condition imposed by the Heisenberg uncertainty principle $0 \leq |\epsilon|\leq \nu$, we see that the fidelity is minimum when
$|\epsilon|=\nu$, so that we get the following lower bound
\begin{equation}
 \mathcal F_{opt}\ge \frac{1+\nu}{1+3 \nu}.
\end{equation}
\end{proof}
Theorems 1 and 2 provide a very useful characterization of the optimal fidelity which can be achieved with Gaussian local operations at Alice
and Bob site. In fact, the bounds are quite tight because the region between the upper and the lower bound is quite small (see
Fig.~\ref{bounds}). Therefore, by simply computing the lowest PT symplectic eigenvalue of the CM of the shared state and using the bounds, one
gets a good estimate of the maximum fidelity that can be obtained with appropriate local operations. In fact, the error provided by the bounds
is never larger than $0.086$ (see Fig.~\ref{error}).

\begin{cor}[\textbf{Upper bound achieved in the symmetric case}]
The upper bound $\mathcal F_{opt}=1/(1+\nu)$ is achieved iff the bipartite Gaussian state shared by Alice and Bob is symmetric. The optimal
local transformation in the symmetric case is a local symplectic map.
\end{cor}
\begin{proof}
The ``if'' part of the theorem directly follows as a special case of the preceding proof. If the Gaussian state shared by Alice and Bob is
symmetric, it is $\det W_a=\det W_b$, implying $\epsilon=0$. Then, Eq.~(\ref{fideps}) shows that in this case the fidelity reaches the upper
bound, $\mathcal F_{opt}=1/(1+\nu)$. Moreover in this case $\theta=\pi/4$ and the local TGCP map of Eqs.~(\ref{tgcpmap}) is optimal and it is a
symplectic one, with $S_a=ZW_aZ \sqrt{2}$, $S_b=W_b\sqrt{2}$, $G_a=G_b=0$. The ``only if'' part instead can be easily proved by using the result
of Theorem 3 about the CM of the optimized bipartite state shown in the following section. The proof is given in the Appendix.
\end{proof}
This latter corollary provides the generalization of the result of Ref.~\cite{ade-illu05}, which obtained the same relation between optimal
fidelity and $\nu$ but by considering only a special class of symmetric Gaussian bipartite state for Alice and Bob, obtained by mixing at a beam
splitter two single-mode thermal squeezed states.

\section{Determination of the optimal local map}

We have derived a lower bound for the optimal fidelity of teleportation of coherent states, by explicitly constructing the family of local TGCP
maps $\omega_{\theta}$ of Eq.~(\ref{tgcpmap}), which transform Alice and Bob shared state so that the corresponding fidelity of teleportation is
given by Eq.~(\ref{fideps}), interpolating between the lower and upper bound of Theorems 1 and 2 by varying $\epsilon = \cos 2\theta$. The map
$\omega_{\theta}$ is symplectic only for $\theta = \pi/4$ and in this case it is the local optimal map for a symmetric shared state, since it
reaches the upper bound of Theorem 1 (see Corollary 1). When $\theta \neq \pi/4$, $\omega_{\theta}$ is a noisy (i.e., non unitary) map, and it
is not optimal in general, because one cannot exclude that different LOCC strategies by Alice and Bob may yield a better noise matrix $N$ and
therefore a larger value of the teleportation fidelity. As discussed in the introduction, here we shall study the optimization of the
teleportation by restricting to GCP maps, which preserves the Gaussian nature of the bipartite state initially shared by Alice and Bob.

In this section we shall derive two results: i) the general form of the final CM of the bipartite Gaussian state after the optimization over all
local GCP maps; ii) the optimal local TGCP map, i.e., the local TGCP map which maximizes the teleportation fidelity when one restricts to TGCP
maps only, excluding in this way measurements of ancillary modes.

Ref.~\cite{fiu02} already provided the analytical procedure for the determination of all the parameters of the optimal TGCP map. Here, by
further elaborating the approach of Ref.~\cite{fiu02}, we will show that the optimal TGCP map can always be written in a simple form, as a local
symplectic operation eventually followed by a single mode \textit{attenuation} \cite{Cav82}, either at Alice or Bob site.

\subsection{Standard form of the correlation matrix of the optimized bipartite state}

In this subsection we show that, even if we do not know the specific form of the optimal GCP map, one can always characterize it indirectly by
determining the general form of its outcome, i.e., the general form of the CM of the final Gaussian state shared by Alice and Bob after the
maximization. We begin with the following lemma.
\begin{lem}[\textbf{Standard form III}] The correlation matrix $V$ of every bipartite Gaussian state can be transformed by
local symplectic operations into the following normal form
\begin{equation}
V_{1}= \left(
\begin{array}{cccc}
n_1  &  &-d_1     & \\
     &n_2   &     &d_2  \\
-d_1     &  &m_1&       \\
     &d_2   &     &m_2
\end{array}
\right),\label{Vsigma}
\end{equation}
where all the coefficients are positive and satisfy the following constraints
\begin{equation}
 n_1-n_2=m_1-m_2=d_1-d_2=\lambda, \qquad \lambda \in \mathbb R \label{constr}.
\end{equation}
That is:
\begin{equation}
V_{1}= \left(
\begin{array}{cccc}
n+\lambda&  &-d-\lambda&    \\
     &n     &     &d    \\
-d-\lambda&     &m+\lambda&     \\
     &d &     &m
\end{array}
\right).\label{Vsigma2}
\end{equation}
\end{lem}
\begin{proof}It is well known that it is possible to transform every $V$ of an entangled state in the usual normal form
(standard form I) \cite{Duan00,Simon00}
\begin{equation}
V_{N}= \left(
\begin{array}{cccc}
a    &  &-c_1     & \\
     &a     &     &c_2  \\
-c_1     &  &b&         \\
     &c_2   &     &b
\end{array}
\right),\label{Vstandard}
\end{equation}
where all the coefficients are positive. Now we perform a local symplectic operation composed by two local squeezing operations,
$S_a=\textrm{diag}(\sqrt{r_a},1/\sqrt{r_a})$ and $S_b=\textrm{diag}(\sqrt{r_b},1/\sqrt{r_b})$. We impose the first two conditions
$n_1-n_2=m_1-m_2=\lambda$,
\begin{equation}
a(r_a-r_a^{-1})=b(r_b-r_b^{-1})=\lambda,\label{nm}
\end{equation}
which solved for positive $r_a$ and $r_b$ give
\begin{eqnarray}
r_a(\lambda)&=&\lambda/(2a)+\sqrt{1+(\lambda/2a)^2}, \label{ra}\\
r_b(\lambda)&=&\lambda/(2b)+\sqrt{1+(\lambda/2b)^2}.\label{rb}
\end{eqnarray}
Now we impose the last constraint $d_1-d_2=\lambda$, that is
\begin{equation}
c_1 \sqrt{r_a(\lambda)r_b(\lambda)}-c_2/ \sqrt{r_a(\lambda)r_b(\lambda)}=\lambda.\label{d}
\end{equation}
Our lemma is proved if there is at least one solution $\lambda$ of Eq.~(\ref{d}). Since $\lambda=0$ is the trivial solution when $c_1=c_2$ and
$V_N=V_1$, we can exclude this particular case and divide Eq.~(\ref{d}) by $\lambda$. Therefore we have to show that the equation
\begin{equation}
f(\lambda)=\frac{1}{\lambda}[c_1 \sqrt{r_a(\lambda)r_b(\lambda)}-c_2/ \sqrt{r_a(\lambda)r_b(\lambda)}]=1\label{equation}
\end{equation}
admits at least one real solution. If $|\lambda|\gg a$ and $|\lambda|\gg b$, we can power expand the square roots in Eqs.~(\ref{ra})-(\ref{rb})
so that we easily find the following limits for $f(\lambda)$:
\begin{eqnarray}
\lim_{\lambda\rightarrow\infty}f(\lambda)&=&c_1/\sqrt{ab}\le1,\label{ine1}\\
\lim_{\lambda\rightarrow-\infty}f(\lambda)&=& c_2/{\sqrt{ab}}\le1,\label{ine2} \\
\lim_{\lambda\rightarrow 0^\pm}f(\lambda)&=& \pm \textrm{sign}(c_1-c_2)\infty.\label{lim0}
\end{eqnarray}
The inequalities in Eqs.~(\ref{ine1})-(\ref{ine2}) follow from the Cauchy-Schwartz inequality $\langle x_1^2\rangle\langle
x_2^2\rangle\ge\langle x_1 x_2\rangle^2$ applied to the quadrature operators of the two modes. Given the three limits (\ref{ine1}), (\ref{ine2})
and (\ref{lim0}), since $f(\lambda)$ is continuous everywhere except at the origin, at least one solution of Eq.~(\ref{equation}) exists.
Moreover this solution has the same sign of $c_1-c_2$.\end{proof} We have defined the standard form of lemma 2 as standard form III because it
is very similar to the standard form II defined in Ref.~\cite{Duan00} for the determination of a necessary and sufficient entanglement criterion
for bipartite Gaussian states. In particular the two standard forms coincide in the special case of a symmetric bipartite state ($n_1=m_1$ and
$n_2=m_2$ or equivalently $n=m$).

\begin{thm}[\textbf{Form of the CM of the optimized bipartite state}]
The optimal GCP map $\omega_{opt}$ maximizing the teleportation fidelity is such that the CM of the transformed bipartite state is in the
standard form III $V_1 $ defined by Eqs.~(\ref{Vsigma})-(\ref{Vsigma2}).
\end{thm}
\begin{proof}
By means of local symplectic operations, we can always put the CM of the bipartite state of Alice and Bob in the form of Eq.~(\ref{Vsigma}), but
without the constraints of Eq.~(\ref{constr}). We first restrict to local symplectic operations and show that the optimal local symplectic
operation always transforms to a state with a CM satisfying the constraints of Eq.~(\ref{constr}). Since the CM is tridiagonal, then any
possible optimal map must be a squeezing transformations of the two modes given by $S_a=\textrm{diag}(r_a,r_a^{-1})$ and
$S_b=\textrm{diag}(r_b,r_b^{-1})$ \cite{fiu02}. Let us define
\begin{eqnarray}
\alpha(r_a,r_b)&=& r_a^2 n_1 -2d_1 r_a r_b+r_b^2 m_1, \\
\beta(r_a,r_b)&=&r_a^{-2}n_2-2d_2 (r_a r_b)^{-1}+r_b^{-2} m_2,
\end{eqnarray}
so that the noise matrix of Eq.~(\ref{noise}) is equal to $N=\textrm{diag}(\alpha,\beta)$. The optimal map must minimize $\det N=\alpha\beta$,
and therefore we impose that
\begin{equation}
 \nabla \alpha(r_a,r_b) \beta(r_a,r_b) =0,\label{ab}
\end{equation}
where $\nabla=(\partial_{r_a},\partial_{r_b})$. Due to Lemma 1, we must also have that $\alpha(r_a,r_b)=\beta(r_a,r_b)\ne 0$, and therefore
Eq.~(\ref{ab}) reduces to
\begin{equation}
 \nabla [\alpha(r_a,r_b)+ \beta(r_a,r_b)] =0\label{a+b}.
\end{equation}
The CM of the transformed state is the optimized one iff the optimal local symplectic operation is the identity map, that is, if
$\alpha(1,1)=\beta(1,1)$ and
\begin{equation}
 \nabla [\alpha(r_a,r_b)+ \beta(r_a,r_b)]\Big|_{r_a=r_b=1} =0.
\end{equation}
It is easy to check that these conditions are satisfied iff
\begin{equation}
 n_1-n_2=m_1-m_2=d_1-d_2,
\end{equation}
which are exactly the constraints of Eq.~(\ref{constr}). The theorem is proved if we show that the normal form $V_1$ of Eq.~(\ref{Vsigma2}) is
actually kept also if one maximizes over the broader class of GCP maps. In fact, if by reductio ad absurdum, we assume that an optimal
(non-symplectic) GCP map exists leading to a CM not satisfying the constraints of Eq.~(\ref{constr}), we could always apply a further symplectic
map which, by repeating the maximization above, would transform to a state with a CM satisfying the constraints (\ref{constr}) and yielding a
larger teleportation fidelity. But this is impossible, because it contradicts the initial assumption of starting from the optimal bipartite
state.
\end{proof}
In other words, the CM of the state shared by Alice and Bob after the maximization of the teleportation fidelity is the one with the standard
form III of Eq.~(\ref{Vsigma2}) because it is the unique CM for which the optimal map is the identity operation on both Alice and Bob site.

\subsection{Optimal trace-preserving Gaussian CP map}

In the former subsection we have determined the form of the CM of the optimized state of Alice and Bob, without determining however which is the
local GCP map which maximizes the teleportation fidelity. Here we find this optimal map, restricting however to the smaller class of
\textit{trace-preserving} GCP maps. The case of Gaussian maps including Gaussian measurements on ancillas will be afforded elsewhere.

Ref.~\cite{lin00} has introduced the notion of \textit{minimal noise} TCGP maps, as the extremal solution of the condition of
Eq.~(\ref{condiTGCP}). These maps are the ones that, for a given matrix $S$, possess the ``smallest'' positive matrix $G$ realizing a CP map. It
is easy to check that a minimal noise TGCP map satisfies the relation $\det G = \left(1-\det S\right)^2$. An example of minimal noise TGCP map
is an \textit{attenuation} \cite{Cav82}, i.e., the transmission of a single boson mode through a beam splitter with transmissivity $\tau$ ($0\leq
\tau\leq 1$), such that
\begin{equation} a \rightarrow \tau a + \sqrt{1-\tau^2} a_V, \label{atte}
\end{equation} where $a=\left(\hat{x}+i\hat{p}\right)/\sqrt{2}$ is the annihilation operator of the mode, and $a_V$ that of the vacuum mode
entering the unused port of the beam splitter.

It is evident that the TGCP map maximizing the teleportation fidelity has to be a minimal noise TGCP map \cite{fiu02}. We prove now a useful
decomposition theorem.
\begin{thm}[\textbf{Decomposition of TGCP maps}]
A minimal noise TGCP map $\omega $ on a single mode system with $\det G\le 1$ can always be decomposed into a symplectic transformation
$\sigma_1$, followed by an attenuation $\tau$ and by a second symplectic transformation $\sigma_2$, that is,
\begin{equation}\label{teo}
\omega =\sigma_2 \circ \tau \circ \sigma_1.
\end{equation}
Therefore a local minimal noise TGCP map on a bipartite CV system can always be decomposed into a local symplectic map, followed by the tensor
product of two local attenuations and by a second local symplectic map. \end{thm}
\begin{proof}
We consider a generic minimal noise TGCP map such that $V \rightarrow V'=SVS^T+G$ for a generic CM $V$, with $\det G = \left(1-\det S\right)^2$.
$G$ is a positive symmetric matrix, and therefore a symplectic matrix $T_{2}$ exists such that $G=T_{2}T_{2}^T (1-s)$, where $s=\det S$. We then
define the symplectic matrix $T_{1}=\frac{1}{\sqrt{s}} T_{2}^{-1} S$, and we also consider an attenuation map with transmissivity $\sqrt{s}$. If
we now first apply the symplectic map defined by $T_1$, then the attenuation map and finally the second symplectic map defined by $T_2$, by
using the relations $T_{2}\sqrt{s}T_{1}=S$ and $G=T_{2}T_{2}^T (1-s)$, one can check that the composition of the three maps reproduces the given
TGCP map.
\end{proof}
\begin{cor} \label{cordec}
A minimal noise TGCP map $\omega $ on a single mode system with $G$ proportional to the identity matrix, i.e., $G=(1-s) I$ $(0 \leq s \leq 1$)
can always be decomposed into a symplectic transformation $\sigma_1$, followed by an attenuation $\tau$,
\begin{equation}\label{teo2}
\omega =\tau \circ \sigma_1.
\end{equation}
Therefore a local minimal noise TGCP map on a bipartite CV system with $G_i=(1-s_i) I$ $(0 \leq s_i \leq 1$, $i=a,b$) can always be decomposed
into a local symplectic map, followed by the tensor product of two local attenuations.
\end{cor}
\begin{proof}It is sufficient to repeat the former proof and consider that, since $G=(1-s) I$, $T_2=I$ and therefore the second symplectic map is
the identity operation.
\end{proof}
This latter case is of interest because the optimal TGCP map must have in fact the property $G_i=(1-s_i) I$, $i=a,b$. To show this we first
simplify the scenario by exploiting the results of Ref.~\cite{fiu02}, which provides the general analytical procedure to derive the optimal
local TGCP map. In fact, Ref.~\cite{fiu02} shows that when the optimal local TGCP map is a minimal noise, non-symplectic one, it can be
performed on one site only, i.e., either on Alice or on Bob alone. Suppose that the non-symplectic map is performed on Bob site; it is
straightforward to see that, under a generic local TGCP map $V \rightarrow V'=SVS^T+G$, with $S=I \oplus S_b$ and $G=I \oplus G_b$, the noise
matrix $N$ transforms according to $N \rightarrow N'=\Gamma+G_b$ where
\begin{equation}\label{transfoN}
\Gamma=Z A  Z  + Z C S_b^T +S_b C^T Z+S_b B S_b^T.
\end{equation}
Ref.~\cite{fiu02} shows that the optimal map is such that $\Gamma \propto G_b$, and since Lemma 1 shows that the optimal $N$ is proportional to
the identity, this implies that $G_b$ must be proportional to the identity. Therefore Corollary 2 leads us to conclude that \textit{the optimal
local TGCP map is either a local symplectic map, or a local symplectic map followed by an attenuation by a beam splitter, placed either on Alice
or on Bob mode}.

Theorem 4 and Corollary 2 therefore provide a very simple and clear description of the TGCP map which maximizes the teleportation fidelity,
which is not evident in the treatment of Ref.~\cite{fiu02}. We can further characterize the optimal local TGCP map by determining: i) the form
of the first local symplectic map; ii) the conditions under which the optimal local operation is noisy, i.e., when one has also to add a beam
splitter with appropriate transmissivity on Alice or Bob mode in order to maximize the teleportation fidelity. In order to do that we first need
a further lemma, similar to lemma 2.

\begin{lem} For any given positive real parameter $\eta$, the correlation matrix $V$ of every bipartite
Gaussian state can be transformed by local symplectic operations into the following normal form
\begin{equation}
V_{\eta}= \left(
\begin{array}{cccc}
n_1  &  &-d_1     & \\
     &n_2   &     &d_2  \\
-d_1     &  &m_1&       \\
     &d_2   &     &m_2
\end{array}
\right),\label{Veta}
\end{equation}
where all the coefficients are positive and satisfy the following constraint
\begin{equation}
 n_1-n_2=\eta(d_1-d_2)=\eta^2(m_1-m_2)=\lambda, \qquad \lambda \in \mathbb R \label{consteta}.
\end{equation}
That is, we have a family of normal forms depending on the parameter $\eta$,
\begin{equation}
V_{\eta}= \left(
\begin{array}{cccc}
n+\lambda   &   &-d-\lambda/\eta    &   \\
        &n  &          &d   \\
-d-\lambda/\eta  &  &m+\lambda/\eta^2  &    \\
        &d  &          &m
\end{array}
\right).\label{Veta2}
\end{equation}
\end{lem}
\begin{proof}
With the same procedure used in the proof of Lemma 2, we arrive to an equation similar to (\ref{equation}), while the corresponding three limits
are exactly the same of (\ref{ine1}), (\ref{ine2}) and (\ref{lim0}), since the factor $\eta$ cancels out. As a consequence the continuity
argument is valid also in this case, and therefore, for every fixed parameter $\eta$, one can find a transformation which puts the CM in the
normal form $V_\eta$.
\end{proof}
We notice two facts that will be useful for the next theorem: i) the optimal CM standard form of Theorem 3, $V_1$,  belongs to the class of
normal forms $V_{\eta}$, since it is obtained for $\eta =1$; ii) when $0 < \eta < 1$, the state with CM $V_{\eta}$ is transformed into the
Gaussian state with CM equal to $V_1$ when a beam splitter with transmissivity $\eta$ is put on Bob mode. We then arrive at the theorem about
the optimal TGCP map.

\begin{thm}
The optimal local TGCP map maximizing the fidelity of teleportation of coherent states can always be decomposed into a local symplectic map,
eventually followed by an attenuation either on Alice or on Bob mode. The first local symplectic map is the one transforming the CM of the
Gaussian state shared by Alice and Bob into one particular normal form of the family $V_\eta$ defined in Eqs.~(\ref{Veta})-(\ref{consteta}),
with $0 < \eta \leq 1$. One has to add the attenuation on one of the two modes for realizing the optimal TGCP map if there is a value of $\eta$,
let us say $\eta=\tau$, such that the coefficients of $V_\tau$ satisfy the relations
\begin{equation}\label{tau}
\tau=\frac{d}{m-1},\qquad \tau<1.
\end{equation}
If instead the condition of Eq.~(\ref{tau}) is never satisfied in the interval $\eta \in [0,1)$, the optimal TGCP map is formed only by the
local symplectic map transforming the CM into the normal form $V_1$ (i.e., $\eta=1$ and no attenuation is required).
\end{thm}
\begin{proof}
Using Lemma 3, we can always apply a local symplectic map which transform the CM into one of the form of the family of Eq.~(\ref{Veta2}), with
$\eta=\tau$. We then apply an attenuation on Bob mode with transmissivity $\tau$ and then try to find the maximum fidelity as in the proof of
Theorem 3. We have that the two diagonal elements of the noise matrix $N$ now read
\begin{equation}
\alpha(\tau)= \beta(\tau)= n-2 \tau d +\tau^2 m+1-\tau^2,
\end{equation}
The optimal map must minimize $\det N=\alpha^2$, and therefore we impose
\begin{equation}
 \frac{d \alpha(\tau)}{d \tau}=0,\label{abeta}
\end{equation}
which is satisfied iff $\tau=d/(m-1)$. If $0 < \tau < 1$, this map composed by the local symplectic map and the attenuation is the optimal map.
If instead for any $\tau \in [0,1]$, the condition of Eq.~(\ref{tau}) is not satisfied, there is no critical point in this interval and
therefore the optimal map is just the symplectic transformation to the normal form $V_1$. One has also to check the behavior at the lower
boundary value $\tau=0$, but this is trivial because this means that Bob uses the vacuum to implement the teleportation which is never the
optimal solution if we have an entangled channel. In fact, if Bob uses the vacuum, the channel looses its quantum nature and the maximum of the
fidelity is the classical one $\mathcal F=1/2$, which is below the lower bound for any entangled state given by Eq.~(\ref{lbound}).
\end{proof}
Theorem 5 therefore characterizes in detail the optimal TGCP map, giving in particular the conditions under which this map is noisy, i.e.,
non-symplectic and therefore when teleportation is improved by \textit{increasing the noise and decreasing the entanglement} of the shared state.

Using this latter theorem we can also determine how, from an operational point of view, one can compute the value of the teleportation fidelity
maximized over all TGCP maps, starting from the symplectic invariants of the bipartite Gaussian state initially shared by Alice and Bob. From
the CM of this latter state one can:
\begin{enumerate}
 \item compute the four symplectic invariants $a=\sqrt{\det A}$, $b=\sqrt{\det B}$, $c=\sqrt{|\det C|}$ and $v=\det V$.
\item Knowing the first three invariants of the channel, the elements $n$, $m$, and $d$ of the normal form $V_{\eta}$ can be expressed as
functions of only the two unknown parameters $\lambda$ and $\eta$ as
\begin{eqnarray}
 n(\lambda) &=& -\lambda/2       + \sqrt{a^2+(\lambda/2)^2},  \\
 m(\lambda, \eta) &=& -\lambda/2\eta^2 + \sqrt{b^2+(\lambda/2\eta^2)^2},  \\
 d(\lambda, \eta) &=& -\lambda/2\eta   + \sqrt{c^2+(\lambda/2\eta)^2}.
\end{eqnarray}
\item The two parameters $\lambda$ and $\eta$ can be found solving the following system in the region $0<\eta<1$,
\begin{equation}\label{sys}
\Bigg \{
 \begin{array}{rcl}
  \det V_{\eta}(\lambda,\eta)&=&v \\
  \eta\; [m(\lambda,\eta)-1]&=&d(\lambda,\eta)\\
 \end{array}
\end{equation}
and solving also the first equation in the boundary $\eta=1$,
\begin{equation}
 \det V_1(\lambda)=v,\label{sys1}
\end{equation}
(the two conditions of (\ref{sys}) come from the invariance property of the determinant of the channel and form the maximization condition of
Eq.~(\ref{tau})).

\item We call $(\lambda_i,\eta_i)$ with $i=1,2\dots k$, the union of the solutions of (\ref{sys}) and
(\ref{sys1}) (we have at least one solution because (\ref{sys1}) admits at least a solution). We then compute the candidate fidelities
\begin{eqnarray}
&& F_i=2\left[2+\sqrt{a^2+\lambda_i^2/4}+\sqrt{b^2\eta_i^4+\lambda_i^2/4}\right. \nonumber \\
&& \left.-2\sqrt{c^2\eta_i^2+\lambda_i^2/4}+1-\eta_i^2 \right ]^{-1}, \label{candifide}
\end{eqnarray}
so that the maximum fidelity will be $\mathcal F_{opt}=\max\{F_i\}$.
\end{enumerate}

\section{Conclusions}

We have studied the Braunstein-Kimble protocol \cite{braKim98} for the CV teleportation of coherent states and how the corresponding fidelity
can be maximized over local operations at Alice and Bob site. We have assumed that Alice and Bob share a Gaussian bipartite state and restricted
to Gaussian LOCC strategies, which preserves such a Gaussian property. We have shown that, for a given shared entanglement, the maximum fidelity
of teleportation is bounded below and above by simple expressions depending upon the lowest PT symplectic eigenvalue $\nu$ only (see
Eq.~(\ref{lowupp})). We have seen that these bounds are quite tight and that the upper bound of the fidelity is reached if and only if Alice and
Bob share a symmetric entangled state. We have also determined the general form of the CM of Alice and Bob state after the optimization
procedure. Then we have restricted to local TGCP maps and shown that the optimal TGCP map is composed by a local symplectic map, eventually
followed by an attenuation either on Alice or on Bob mode. Finally we have shown how the corresponding value of the maximum fidelity
$\mathcal{F}_{opt}$ can be derived from the knowledge of the symplectic invariants of the initial CV entangled state shared by Alice and Bob.

If one considers generic GCP maps (i.e., including also Gaussian measurements on ancillas), one expects to further improve the teleportation; in
this case one should adopt the description of GCP maps given in \cite{giedkemap} in order to characterize the optimal local Gaussian map, but
this will be the subject of future work.

Another open question is to see if teleportation fidelity can be increased by leaving the Gaussian setting studied here and consider more
general non-Gaussian local operations at Alice and Bob site. In this respect, the preliminary results of Refs.~\cite{nonGtelep} seem promising.

\section{Acknowledgements}

This work has been supported by the European Commission through the FP6-IP QAP.

\section{Appendix}

We now prove the ``only if'' part of Corollary 1, i.e., that if $\mathcal F_{opt}=1/(1+\nu)$ (the upper bound of the optimal fidelity), then the
bipartite Gaussian state shared by Alice and Bob is symmetric.
\begin{proof}
Theorem 3 shows that the final CM after any optimization map must be $V_1$ of Eq.~(\ref{Vsigma2}). Using Lemma 1 and the explicit form of $V_1$,
the hypothesis is equivalent to $n+m-2d=2\nu$, so we can make the substitution $d=(n+m)/2-\nu$, which is just a different parametrization:
$V_1(n,m,d,\lambda)\rightarrow V_1(n,m,\nu,\lambda)$. Now, the condition that $\nu$ is equal to the PT minimum symplectic eigenvalue gives us a
constraint on the parameters of the matrix $V_1$
\begin{equation}
\nu\{V_1(n,m,\nu,\lambda)\}=\nu. \label{nuconstr}
\end{equation}
If the state is symmetric, which means $n=m$, then the condition of Eq.~($\ref{nuconstr}$) is identically satisfied. If the state is non
symmetric, Eq.~(\ref{nuconstr}) is a non-trivial equation that solved for $\lambda$ gives $\bar \lambda=(m-n)^2/8\nu-n-m$. However the
corresponding matrix $V_1=(n,m,\nu,\bar \lambda)$ is not the CM of a physical state; in fact, the characteristic polynomial of $V_1$ can be
written as $P(x)=(c_0+c_1 x + x^2)(g_0+g_1x+x^2)$ where $c_0=-\nu(n+m+\nu)$, but this means that $V_1$ has at least one negative eigenvalue and
therefore it is not positive definite. Therefore $\mathcal F_{opt}=1/(1+\nu)$ is realized only if Alice and Bob state is a Gaussian symmetric
state.
\end{proof}


\begin{thebibliography}{99}
\bibitem{notedispl}If the state shared by Alice and Bob has zero displacement and Bob always performs the ``correct''
conditional displacement, the teleportation protocol is invariant under displacement transformation, i.e., all states with the same covariance
matrix but different coherent components are teleported with the same fidelity \cite{fiu02,pirrew}.
\bibitem{noterefe} This impossibility to improve entanglement in entanglement swapping protocols is always valid for CV systems, even
under postselection, which affects only first-order moments of the swapped state. In the case of qubits instead, this impossibility holds only
for the ensemble fidelity, averaged over all the Bell measurements results, because it is possible to have Bell measurement results which
probabilistically improve the fidelity.
\end{thebibliography}
\end{document}